\documentclass[twocolumn,showpacs,preprintnumbers,amsmath,amssymb]{revtex4}

\usepackage{graphicx}
\usepackage{dcolumn}
\usepackage{bm}

\begin{document}

\title{Emergence of fractal behavior in condensation-driven aggregation}%

\author{M. K. Hassan\footnote{Md. Kamrul Hassan, Electronic address: 
khassan@univdhaka.edu}$^a$ 
and M. Z. Hassan\footnote{Md. Zahedul Hassan, Electronic address: ${\rm zahed_-aec}$@yahoo.com}$^b$
}%
\date{\today}%

\affiliation{
$a$  Theoretical Physics Group, Department of Physics, University of Dhaka, Dhaka 1000, Bangladesh \\
$b$ ICT Cell, Bangladesh Atomic Energy Commission, Dhaka 1000, Bangladesh 
}

\begin{abstract}%
We investigate a model in which an ensemble of chemically identical Brownian particles are 
continuously growing by condensation and at the same time undergo irreversible aggregation whenever two particles come 
into contact upon collision. We solved the model exactly by using scaling theory for the case whereby a particle, say 
of size $x$, grows by an amount $\alpha x$ over the time it takes to collide with another particle of any size.
It is shown that the particle size spectra of such system exhibit transition to dynamic scaling 
$c(x,t)\sim t^{-\beta}\phi(x/t^z)$ accompanied by the emergence of fractal of dimension 
$d_f={{1}\over{1+2\alpha}}$. One of the remarkable feature of this model is that it is governed by a non-trivial conservation law,
namely, the $d_f^{th}$ moment of $c(x,t)$ is time invariant regardless of the choice of the initial conditions. 
The reason why it remains conserved is explained by using a simple dimensional analysis. 
We show that the scaling exponents $\beta$ and $z$ are locked with the fractal dimension $d_f$ via a 
generalized scaling relation $\beta=(1+d_f)z$. 
\end{abstract}

\pacs{61.43.Hv, 64.60.Ht, 68.03.Fg, 82.70Dd}

\maketitle

\section{Introduction}

The formation of clusters by aggregation of particle, its underlying causes and consequences, 
is one of the most fundamental yet challenging problem of many 
processes in physics, chemistry, biology and engineering. 
Examples include aggregation of colloidal or aerosol particles suspended in liquid or gas 
\cite{ref.friedlander, ref.thorn,ref.melle}, polymerization \cite{ref.polymerization}, 
antigen-antibody aggregation \cite{ref.antigen} and cluster formation in galaxy \cite{ref.galaxy}. 
Such a wide variety of applications has resulted in numerous studies focusing mostly on the kinetic 
and geometric aspects of the problem. The kinetic aspect is well studied and well understood by theory, 
experiment and by numerical simulation. The first successful theoretical model was proposed 
more than one hundred years ago by von Smoluchowski and it still remains the only analytical 
model which has provided much of our theoretical understanding \cite{ref.smoluchowski}. The definition of this model is trivially simple.
It is assumed that initially an ensemble of chemically identical particles undergo sequential 
aggregation upon collision. 

The kinetics of aggregation by the Smoluchowski equation has been extensively studied in and around the $1980$s and 
significant contribution was made during this period especially on the scaling theory and gelation transition \cite{ref.ziff,ref.scaling,ref.leyvraz}.
One of the most striking results is that the concentration $c(x,t)$ of particles of size $x$ at time $t$ 
exhibits dynamic scaling 
\begin{equation}
\label{eq:0}
c(x,t)\sim s(t)^{-\theta}\phi\Big (x/s(t)\Big ),
\end{equation}
in the limit $t\rightarrow \infty$, where $\phi(\xi)$ is the scaling function, $s(t)$ is the mean particle size
and the conservation of mass principle tunes the mass exponent to an integer value $\theta=2$ \cite{ref.vicsek, ref.scaling}. 
The structure of the above scaling ansatz is highly instructive as it has been found in many seemingly unrelated phenomena.
It implies that there must exists a common underlying 
mechanism for which such disparate systems behave in a remarkably similar fashion \cite{ref.scaleinvariance}.
On the other hand, the insights into the geometric
aspect is mostly provided by experiments and numerical simulations and these studies reveal that
when particles aggregate almost always scale-invariant fractals emerge \cite{ref.vicsek}.  
Unfortunately, there does not yet exist an analytically solvable model which could help us know why fractals are 
ubiquitous in the aggregation process.

In addition to aggregation, particles may also grow in size by condensation, deposition or by 
accretion. For instance, aggregation in vapor phase or in damp 
environment particles or droplets may continuously grow by condensation 
\cite{ref.droplet,ref.sire,ref.husar,ref.dust}. 
It is also well-known that aerosol or colloidal particles are often not stable rather their sizes may evolve via 
aggregation and condensation leading to gas-to-particle conversion. 
However, in the absence of impurity such as dirt or mist, the condensation can only take place on the existing 
particles without forming new nuclei provided the concentration of particles present is 
sufficiently high and the supersaturation is sufficiently low \cite{ref.friedlander,ref.husar}. This type of growth
is known as the heterogeneous condensation.
To this end, we recently proposed a simple condensation-driven aggregation (CDA) model
and discussed the kinetic aspect of the problem  \cite{ref.pre}. 
In this article, we present an alternative method to solve the CDA model and kept our focus 
mainly on its geometric aspects instead. We show analytically that the resulting system can be best described as fractal 
and quantified by its dimension $d_f$ which decreases with increasing strength of growth by condensation. 
Interestingly, we find that the key results of the CDA model are connected, in one way or another, 
to the fractal dimension $d_f$. For instance, the $d_f$th moment of the distribution function $c(x,t)$
is a conserved quantity, the mean particle size grows with time as $t^{{{1}\over{d_f}}}$,
in terms $d_f$ we can write a generalized scaling relation $\beta=(1+d_f)z$, etc. 
To test our analytical predictions, we have performed extensive numerical simulation and 
found that analytical results are in perfect agreement with numerical data.

The rest of the paper is organized as follows. In Sec. II, we give a detailed description of the 
CDA model including its algorithm.
In Sec. III, some of the key features of the model are discussed.
In Sec. IV, we give a simple dimensional analysis to the governing equation of the CDA model in an attempt to gain
deeper insight into the scaling theory. We applied the scaling theory in Sec. V to obtain the solution for
the distribution function $c(x,t)$. In Sec. VI, we invoke the idea of fractal analysis to the CDA model.
Finally, in Sec. VII we discuss and summarize our work. 

\section{The Model}

Chemically identical particles in aggregation process are typically characterized by their mass or size and shape. However, 
if the particles are one dimensional then size or mass is the only dynamical variable. Therefore, 
within a given class of units of measurement both mass and size 
can be described by the same numerical value as they differ only by a proportionality constant. However, this
is not true in the case of higher dimensional particles. 
The Smoluchowski model is inherently one dimensional 
and hence size and mass can be used interchangeably. In the CDA model, we characterize each particle
by the size it assumes upon aggregation till it takes part in further aggregation.
The extent of growth by condensation can be quantified by the
growth velocity defined as the ratio of the net growth and the elapsed time during which 
this growth occurs. The most natural choice for the elapsed time 
is definitely the collision time. 
The growth velocity is then fully specified if we know the amount of growth 
of a given particle which occurs between collisions. For this, we assume that the net growth of a particle 
between collisions, in the most generic case, is directly proportional to the size by which it is characterized. 
That is, a particle which is just born upon aggregation with size $x$ will have its
size equal to $x+\alpha x$ whenever it collides with another particle regardless of the amount of time it takes to collide.
This is not at all a bad assumption since such a choice makes the growth velocity 
stochastic in nature as the growth size and the collision time both become random in 
character.  

\begin{figure} 
\includegraphics[width=8.50cm,height=4.15cm,clip=true]{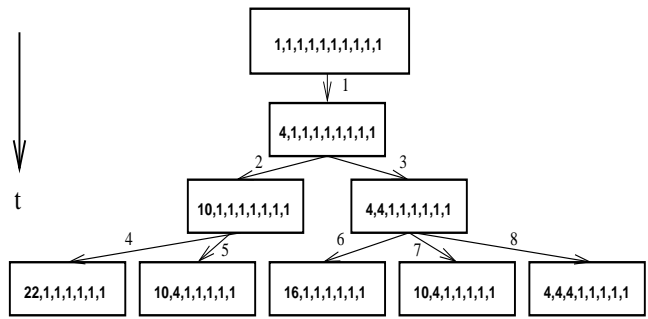} 
\caption{Schematic representation of the model for $\alpha=1$ is given by
using monodisperse initial condition as an example.
}                                                
\label{fig1}
\end{figure}
For numerical simulation, one may think of keeping a logbook where the sizes of 
the particles are registered each time they take part in aggregation.
Initially, sizes of all the particles in the system are registered in the logbook.  The rules
these particles then have to follow at each step during simulation are:
\begin{itemize}
\item[{\bf(i)}] Two particles are picked randomly from the system to mimic random collision via Brownian motion.
\item[{\bf (ii)}] The sizes of the two particles are increased by a fraction $\alpha$ 
of their respective sizes in the logbook to mimic the growth by condensation. 
\item[{\bf (iii)}] Their sizes are combined to form one particle to mimic the aggregation process. 
\item[{\bf (iv)}] The logbook is updated by registering the size of the new particle 
in it and at the same time deleting the sizes of its constituents from it.
\item[{\bf (v)}] The steps (i)-(iv) are repeated {\it ad infinitum} to mimic the time evolution. 
\end{itemize}
In order to illustrate how these rules of the model work for 
monodisperse initial condition, we give a simple 
example in Fig. $(1)$ using an  evolutionary tree based approach. 

The CDA model can also be understood by a reaction scheme written as  
\begin{equation}
\label{eq:reaction} A_x(t) + A_y(t) \stackrel{
v(x,t)}{\longrightarrow} A_{(\alpha + 1)(x + y)}(t + \tau),
\end{equation}
where $A_x(t)$ denotes the aggregate of size $x$ at time $t$ and $\tau$ is the elapsed time. 
This reaction scheme can be described by the following generalized Smoluchowski (GS) equation
\begin{eqnarray}
\label{eq:1}
\Big[{{\partial }\over{\partial t}} &+&  {{\partial}\over{\partial x}} v(x,t) \Big]c(x,t)
=-c(x,t)\int_0^\infty K(x,y)c(y,t)dy \nonumber \\ &+ & {{1}\over{2}}\int_0^x dy K(y,x-y)
c(y,t)c(x-y,t).
\end{eqnarray}
The second term on the left hand side of the above equation accounts for the growth by condensation with
velocity $v(x,t)$. On the other hand, the first (second)
term on the right hand side of Eq. (\ref{eq:1}) describes the loss (gain) of size $x$ 
due to merging of size $x$ ($(x-y)$) with particle of size $y$.
However, the GS equation can only describe the CDA model if the growth velocity $v(x,t)$, 
the collision time $\tau$, and the kernel $K(x,y)$ are suitably chosen as required by the rules (i)-(v).  
For instance, according to rule (ii) of our model, the net growth of a particle 
of size $x$ between collisions is $\alpha x$. To obtain a suitable expression for the elapsed time we do a 
simple dimensional analysis in Eq. (\ref{eq:1}) and immediately find that 
the inverse of $\int_0^\infty K(x,y)c(y,t)dy$ is the collision time $\tau(x)$
during which the growth $\alpha x$ takes place \cite{ref.maslov}. 
The mean growth velocity between collisions therefore is
\begin{equation}
\label{eq:2}
 v(x,t) = {{\alpha x}\over{\tau(x)}}=\alpha x\int_0^\infty dyK(x,y)c(y,t).
\end{equation} 
The rule $(i)$ on the other hand says that a given particle can 
collide with any particle in the system with an equal {\it a priori} probability regardless of their size. 
This can be ensured only if we choose an aggregation kernel independent of its argument 
and hence we set
\begin{equation}
\label{eq:3}
 K(x,y) = 2,
\end{equation}
for convenience.

\section{Some of the basic features}

\begin{figure}
\includegraphics[width=8.50cm,height=4.25cm,clip=true]{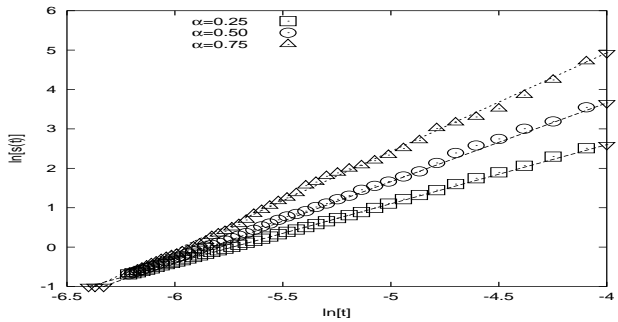}
\caption{ Plots of $\ln(s)$ versus $\ln(t)$ are shown for three different $\alpha$ values but w
ith
the same monodisperse initial conditions in each case.
The lines have slopes equal to $(1+2\alpha)$ revealing the same growth law as predicted by Eq.
(\ref{eq:s}).
}
\label{fig2}
\end{figure}

\begin{figure}
\includegraphics[width=8.50cm,height=4.25cm,clip=true]{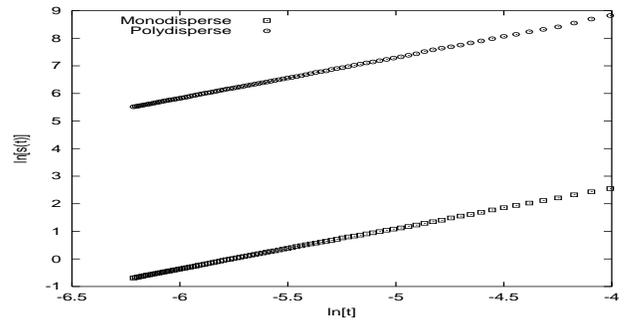}
\caption{Plots of $\ln (s)$ versus $\ln (t)$ for monodisperse (all the particles are chosen to
be of unit size) and polydisperse (e.g., the numerical value for the size of the $500$ 
particles are chosen randomly from the interval $[1,10,000]$) initial conditions. 
Two parallel lines prove that the growth-law for $s(t)$ given by Eq. (\ref{eq:s}) is 
independent of initial conditions.
}
\label{fig3}
\end{figure}

It is noteworthy to mention that the distribution function $c(x,t)$ itself is not a directly observable quantity
but its various moments are. Therefore, one often finds it more convenient to deal with its moment than
the function itself.  The $k$th moment of $c(x,t)$ is defined as
\begin{equation}
\label{eq:4}
M_k(t)=\int_0^\infty x^kc(x,t)dx, \hspace{0.30cm} {\rm with} \hspace{0.25cm} k \geq 0.
\end{equation}
Incorporating it in Eq. (\ref{eq:1})
after substituting Eqs. (\ref{eq:2}) and (\ref{eq:3}) in it we obtain
\begin{eqnarray}
\label{eq:5}
{{dM_k(t)}\over{dt}} & = & \int_0^\infty\int_0^\infty dx dy c(x,t)c(y,t) \\ \nonumber
& \times & \Big [(x+y)^k+(\alpha k-1)(x^k+y^k)\Big].
\end{eqnarray}
In the case of $\alpha=0$ which describes the classical Smoluchowski (CS) equation 
we find ${{dM_1(t)}\over{dt}}=0$ and hence $M_1(t)=\int_0^\infty xc(x,t)dx$ is a conserved quantity . This is well 
known as the conservation of mass principle. Obviously, this principle is no longer obeyed in the CDA model because of
the growth by heterogeneous condensation. 
To check we solve Eq. (\ref{eq:5}) for $M_1(t)\equiv  {\cal L}(t)$ and find that
\begin{equation}
\label{eq:10}
{\cal L}(t)\sim t^{2\alpha}.
\end{equation}
in the long time limit. Hence, the growth of the total mass or size exhibits non-universal behaviour in the sense that its
exponent depends on the parameter $\alpha$. It confirms that the conservation of mass principle is
violated $\forall \alpha>0$. 
Then the question remains: Can there be another conservation law which the system should obey as it evolves? 
One cannot find a straightforward 
answer to this question from Eq. (\ref{eq:5}) except the $\alpha=0$ case.  
 We now solve Eq. (\ref{eq:5})
for $k=0$ to obtain the solution for the total number of particles $M_0(t)\equiv N(t)$ 
present at time $t$ and find that 
in the long time limit it decays algebraically with a universal exponent  
\begin{equation}
N(t) \sim t^{-1}.
\end{equation}
In other words, the number density $N$, evolves following the same differential 
equation as the one we would obtain for the CS equation. It therefore confirms that condensation takes place 
only on the existing particles without forming new nuclei.

Using the solutions for the first two moments in the definition for the mean particle size 
$s(t)= {\cal L}(t)/N(t)$ we obtain the following growth-law
\begin{equation}
\label{eq:s}
s(t)\sim t^{1+2\alpha},
\end{equation}
in the long time limit. 
To verify this, we performed numerical simulation based on 
the rules (i)-(v). However, in order to manipulate the numerical data
we define time $t=1/N$ since the number of particles present in 
the system determines how fast or slowly the aggregation process should proceed.
We first use the monodisperse initial condition where all the particles are assumed to be of unit size.
In Fig. (2), we present plots of $\ln (s)$ 
versus $\ln (t)$ from the resulting data and find three straight lines for three different values of 
$\alpha$. The slopes of these lines satisfy the relation $z=1+2\alpha$, $\forall \ \alpha>0$ which 
clearly shows algebraic growth of $s(t)$ as predicted by Eq. (\ref{eq:s}). The next important thing is to check if 
the initial distribution of particle size has any effect in this growth law.  To find this out we 
simulated the model for several different polydisperse initial conditions and collected data for the mean
particle size $s(t)$ against time $t$. In one of the instances, we picked initially $500$ particles of size chosen 
randomly from the interval $[1,10,000]$ and let the program run in the computer following the rules (i)-(v) 
of the algorithm. In Fig. (3), we again present plots of $\ln (s)$ versus $\ln (t)$ from the resulting numerical data and put 
it together with the corresponding plot for the monodisperse initial 
condition to see the contrast. Surprisingly, we find two parallel lines which clearly implies that the exponent of the 
growth law is universal in the sense that it is independent of the initial conditions. 
That is, one may choose any number of his choice to replace the particles characterized by $1$ in the first box ($t=0$) of 
Fig. (1) and let them play following the rules (i)-(v). One would still obtain the same growth-law for the  
mean particle size as the one for the monodisperse initial condition.

\section{Dimensional Analysis}

There are two governing parameters $x$ and $t$ in the GS equation. However, according to Eq. (\ref{eq:s})
the size of the particle can be expressed in terms of time.
Therefore, only one of the variable, say, time $t$ can be taken as an independent parameter. The other
governing parameter such as $x$ and the governed parameter $c(x,t)$ both can be expressed as a function of 
$t$ alone. We already know from Eq. (\ref{eq:s}) that $t^z$ with kinetic exponent 
\begin{equation}
\label{kinetic}
z=1+2\alpha,
\end{equation}
have the dimension of length $[s(t)]=L$ and hence we can define a dimensionless quantity 
\begin{equation}
\label{eq:xi}
\xi={{x}\over{s(t)}}.
\end{equation} 
On the other hand, applying the
power-monomial law for the dimension of physical quantity we can write a dimensional relation 
$c(x,t)\sim t^{-\beta}$ where the exponent $\beta$ assumes such a value that makes
$t^{-\beta}$ bear the dimension of $c(x,t)$ \cite{ref.barenblatt}. We can therefore define 
yet another dimensionless quantity $\phi$ as follows
\begin{equation}
\label{eq:governed}
\phi=\frac{c(x,t)}{t^{-\beta}}.
\end{equation} 
Now, within a given class one can pass from one units of measurement to another system of units of measurement 
by changing $t$ by an arbitrary factor leaving the other factor unchanged. Upon such a transition the quantity on the 
right hand side of Eq. (\ref{eq:governed}) remain unchanged
since the left hand side is a dimensionless quantity. 
It means that the quantity $\frac{c(x,t)}{t^{-\beta}}$ {\it vis-a-vis} $\phi$ can only be a function of
another dimensionless quantity and the only dimensionless governing parameter is $\xi$ given by Eq. (\ref{eq:xi}).
We therefore find that Eq. (\ref{eq:governed}) leads to the following scaling ansatz  
\begin{equation}
\label{eq:ansatz1}
c(x,t)\sim t^{-\beta}\phi(x/t^z),
\end{equation}
or can be expressed in the form of Eq. (\ref{eq:0}) if we set $\beta=\theta z$ and use 
Eq. (\ref{eq:s}) thereafter. 
Existence of scaling means the following. The quantity $c(x,t)$ that depends on two variables $x$ and $t$
is considered to admit scaling if the two variables combine into one variable 
such away that it can be expressed as Eq. (\ref{eq:ansatz1}).
The fact that two variables combine into one variable 
leads to an enormous simplification in finding the solution to the problem as we shall see below.

\section{Scaling solution}
  
To check if the solution of the GS equation exhibits scaling or not, we substitute Eq. (\ref{eq:ansatz1}) together
with $\beta=\theta z$ in the GS 
equation after substituting Eqs. (\ref{eq:2}) and (\ref{eq:3}) in it and obtain
\begin{eqnarray}
\label{eq:8}
t^{\theta z-z-1} & = & {{1}\over{F(\xi)}}{\Big [}(2(1+\alpha)\Phi_0\phi(\xi) 
+2\alpha\Phi_0 \xi{{d\phi}\over{d\xi}}
\nonumber \\ & - & \int_0^\xi \phi(\eta)\phi(\xi-\eta)d\eta{\Big ]},
\end{eqnarray}
where, 
\begin{equation}
F(\xi)= \theta z\phi(\xi)+z\xi \phi^\prime(\xi),
\end{equation}
and $\Phi_0=\int_0^\infty \phi(\xi)d\xi$ is the zeroth moment of the scaling function 
$\phi(\xi)$. Note that the right hand side of Eq. (\ref{eq:8}) is dimensionless while the left hand side is not.
Thus the dimensional consistency requires
\begin{equation}    
\label{eq:z}
z={{1}\over{\theta-1}}.
\end{equation}
Substituting the $z$ value from Eq. (\ref{kinetic}) in the above equation we obtain the mass exponent 
\begin{equation}
\label{eq:theta}
\theta=\frac{2+2\alpha}{1+2\alpha}.
\end{equation}
The values for the exponents $\theta$ and $z$ are exactly the same as obtained by the exactly solvable method, namely
the Laplace transformation and the method of characteristics, in 
Ref. \cite{ref.pre}. 
To obtain the complete scaling or self-similar solution of the GS equation we still have 
to find $\phi(\xi)$. For that,
we substitute the value of $z$ and $\theta$ in Eq. (\ref{eq:8}) to get
\begin{eqnarray}
\label{eq:ordinary}
& & \Big [1+2\alpha(1-\Phi_0)\Big ]\xi\frac{d\phi}{d\xi}+ \int_0^\xi\phi(\eta)\phi(\xi-\eta)
d\eta \nonumber \\ 
&+ & \Big [2(1+\alpha)(1-\Phi_0)\Big ]\phi(\xi)=0
\end{eqnarray}
The solution of the problem thus reduces to finding the solution of an ordinary 
integro-differential 
equation for the scaling function $\phi(\xi)$.

We now multiply on both sides of Eq. (\ref{eq:ordinary}) by $\xi^n$ and 
integrate from $\xi=0$ to $\xi=\infty$ to obtain an equation for the $n$th moment $\Phi_n$
of $\phi(\xi)$ (or Mellin transform $\Phi(n+1)=\Phi_n)$  
which can be written in the closed form
\begin{equation}
\label{eq:phi}
2\alpha n(1-\Phi_0)\Phi_n+(n-1)\Phi_n+2\Phi_n\Phi_0=\sum_{r=0}^n ~^nC_r\Phi_r\Phi_{n-r},
\end{equation}
for integer value of $n$ only. By setting $n=0$ in the above equation we find
\begin{equation}
\Phi_0(\Phi_0-1)=0,
\end{equation}
and the only non-trivial solution of this equation is $\Phi_0=1$. Using it 
back in Eq. (\ref{eq:ordinary}) gives
\begin{equation}
\label{eq:scaling}
\xi{{d\phi}\over{d\xi}}=- \int_0^\xi\phi(\eta)\phi(\xi-\eta)d\eta.
\end{equation}
To find a solution of this equation we set $\Phi_0=1$ in Eq. (\ref{eq:phi}) and find
the following hierarchy of equations for different integer $n$ values and a few of these are
\begin{eqnarray}
\label{eq:hierarchy}
& & \Phi_2=2\Phi_1^2; \hspace{0.25cm} \Phi_3=3\Phi_2\Phi_1; \\ \nonumber & & 
\Phi_4={{1}\over{3}}\big [ 8\Phi_3\Phi_1+6\Phi_2^2\big ];\hspace{0.25cm}  ...\ ... \hspace{0.15cm}  {\rm e.t.c.,}.
\end{eqnarray}
The solution $\Phi_0=1$ and its definition 
\begin{equation}
\Phi_0=\int_0^\infty \xi^0\phi(\xi)d\xi,
\end{equation} 
implies 
\begin{equation}
\label{eq:scalingsolution}
\phi(\xi)=e^{-\xi},
\end{equation}
which is in fact the inverse Mellin transform of $\Phi(1)=\Phi_0$. It can also 
be verified by substituting it in Eq. (\ref{eq:scaling}). 
Indeed, one can check that this solution does satisfy the hierarchy of all the relations in Eq. (\ref{eq:hierarchy}) and 
simultaneously it solves Eq. (\ref{eq:scaling}). 

Substituting Eqs. (\ref{eq:s}), (\ref{eq:theta}) and (\ref{eq:scalingsolution}) in 
Eq. (\ref{eq:0}) we can finally obtain the scaling solution of the GS equation
\begin{equation}
\label{eq:solution}
c(x,t)\sim t^{-(2+2\alpha)}e^{-{{x}\over{t^{1+2\alpha}}}}.
\end{equation}
This is exactly what we found in Ref. \cite{ref.pre} from the explicit time dependent solution  
by using the limit $t\longrightarrow \infty$. However, the exact solution in \cite{ref.pre} was obtained only for the
monodisperse initial condition. The advantage of using the scaling theory is that we do not need 
to specify the initial condition. It implies that the various essential features or laws of the
CDA model should remain independent of the initial condition
provided the number of particles present at time $t=0$ is sufficiently large and one allows the
system to run for sufficiently long time.  
We can now obtain the solution of Eq. (\ref{eq:5}), the $k$th moment of $c(x,t)$ by 
using Eq. (\ref{eq:solution}) in its definition given by Eq. (\ref{eq:4}) to give
\begin{equation}
\label{eq:20}
M_k(t) \sim t^{z\Big (k-{{1}\over{1+2\alpha}}\Big )}.
\end{equation}
A surprising feature of this solution is that it implies the existence of 
a non-trivial conservation law namely the $q$th moment is independent 
of time if $q={{1}\over{1+2\alpha}}$. In numerical simulation, it means that the sum of the 
$\Big ({{1}\over{1+2\alpha}}\Big )$th power of all the particles is a constant as a function of time.
To check this, we plot $\ln (M_q)$ with $q={{1}\over{1+2\alpha}}$ against 
$\ln (t)$ from the numerical data for different $\alpha$ values and for different
initial conditions. We find a set of parallel horizontal lines for all $\alpha>0$ values
(see Fig. (4)) regardless of initial conditions  value revealing that our analytical result 
is in perfect agreement with the numerical simulation.
One may ask: What is so special about this $q$ value that makes this moment a conserved
quantity? To find an answer to this question we invoke the idea of fractal analysis which is discussed below.

\begin{figure}
\includegraphics[width=8.50cm,height=4.25cm,clip=true]{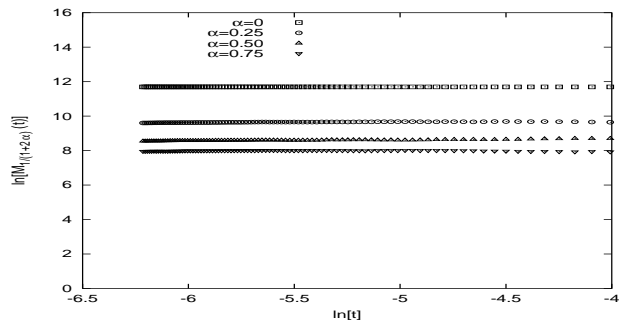}
\caption{Plots of the $\Big ({{1}\over{1+2\alpha}}\Big )$th moment of the particle size 
distribution function versus time for different initial conditions 
(monodisperse and polydisperse) and for different $\alpha$ values are drawn using numerical data. 
A set of horizontal lines clearly prove that the generalized conservation law is always obeyed regardless of the
initial conditions or the $\alpha$ value.
}
\label{fig4}
\end{figure}

\section{Fractal analysis of the CDA model}

In this section, we intend to address the geometric aspect of the CDA model by invoking the idea of fractal analysis.
Before doing so we find it worthwhile to appreciate the following. 
Theoretically, particles in the CDA model can be considered to be inside a row of boxes forming a $1$d lattice with each box as one 
lattice point. Initially, each of these boxes contains one particle characterized by one number whose size distribution
depends on the initial particle size distribution. In this sense, particles are in fact embedded in a space of 
dimension equal to one. The geometry of the resulting system therefore will be called fractal if the dimension of the distribution of particles 
is less than one and greater than zero. To know exactly what this value is
we define ${\cal L}(t)$ as the measure which is the sum of all the aggregates in different boxes at  
time $t$ and obviously it is an ever growing quantity against $t$ according to Eq. (\ref{eq:10}).

To quantify the measure ${\cal L}$, one can use a suitable  
yardstick and find an integer number $N$ needed to cover the measure ${\cal L}$. The most suitable candidate for the 
yardstick in the context of the CDA model is the mean particle size
$s(t)$ which will always give the number $N$ an integer value. 
That is, the size of the measure ${\cal L}(t)$ can be quantified by the number $N(s)$.
Using $k=0$ in Eq. (\ref{eq:20}), we can easily see that the number $N(s)$, when expressed in terms of $s(t)$, exhibits power-law
\begin{equation}
\label{eq:29}
N(s)\sim s^{-d_f},
\end{equation}
with exponent
\begin{equation}
\label{eq:30}
d_f={{1}\over{1+2\alpha}},
\end{equation}
which is highly significant for the following reason. 
Note that when the number $N$ is obtained by measuring a given measure with a suitable yardstick and find that 
it exhibits power-law against the size of the yardstick, then the exponent of the power-law is widely known as
the Hausdorff-Besicovitch (H-B) dimension \cite{ref.feder}. On the other hand, the H-B dimension is called fractal if it is non-integer 
and at the same time if it is less than the dimension of the embedding space. 
It implies that the exponent $d_f$ of Eq. (\ref{eq:29}) is the fractal dimension of the resulting system 
since, according to Eq. (\ref{eq:30}), $d_f$ is not only non-integer $\forall \ \alpha>0$ but also less than the dimension of the
embedding space $d=1$. It is noteworthy that the size of the fractal that emerges in the CDA model is continuously growing  
with time but at the same time it preserves its dimension which is ensured by the conservation law.
Within the rate equation approach, such fractal analysis was first done by Ben-Naim and Krapivsky in the context of 
the stochastic Cantor set \cite{ref.fractal} and later one of us applied it successfully in 
several different systems \cite{ref.fractal_1}.
To verify our analytical result, we have drawn $\ln (N)$ versus $\ln (s)$ in Fig. (5) from the numerical data collected
for a fixed initial condition but varying only the
$\alpha$ value. On the other hand, in Fig. (6) we have drawn the same plots for a fixed $\alpha$ value but varying only
initial conditions (monodisperse and polydisperse). Both figures show an excellent
power-law fit as predicted by Eq. (\ref{eq:29}) with exponent exactly equal to $d_f$ regardless of the choice we make
for the initial size distribution of particles in the system. 

\begin{figure}
\includegraphics[width=8.50cm,height=4.25cm,clip=true]{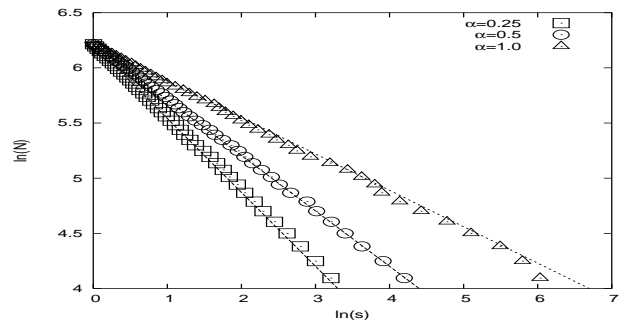}
\caption{Plots of $\ln(N)$ vs $\ln(s)$ are drawn for three different $\alpha$
values keeping the same initial condition. The lines have slopes equal to
${{1}\over{1+2\alpha}}$ which is exactly what was predicted by the theory.
}
\label{fig5}
\end{figure}
\begin{figure}
\includegraphics[width=8.50cm,height=4.25cm,clip=true]{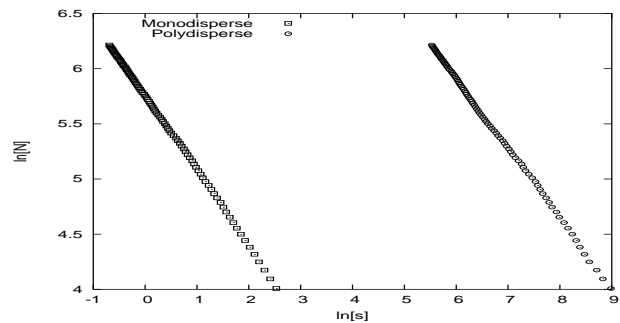}
\caption{Two parallel lines resulting from the plots of  $\ln (N)$ versus $\ln(s)$
for monodisperse and polydisperse initial condition reveal that $N\sim s^{-d_f}$ is
independent of initial conditions.
}
\label{fig6}
\end{figure}
We shall now show that the various interesting results of the CDA model can be expressed in terms
of the fractal dimension $d_f$. For instance, we can use the expression for the fractal dimension $d_f$ in 
Eq. (\ref{kinetic}) to obtain $z={{1}\over{d_f}}$. Using it in Eq. (\ref{eq:s})  
we obtain the following growth law for the mean particle size
\begin{equation} 
s(t)\sim t^{{{1}\over{d_f}}}. 
\end{equation}
We find it worthwhile to mention here, as a passing note, that a similar growth
law has also been found experimentally by Weitz {\it et al} while studying the diffusion-limited 
cluster-cluster aggregation \cite{ref.weitz}. 
Also, the mass exponent $\theta$ can be expressed in terms of $d_f$ 
by using Eq. (\ref{eq:30}) in Eq. (\ref{eq:theta}) to give 
\begin{equation}
\label{eq:massexponent}
\theta = 1+d_f.
\end{equation}
We thus find that $\theta$ always satisfies the inequality $\theta<2$, $\forall \ \alpha>0$ and the inequality
becomes equality $\theta=2$ only if $\alpha=0$ which corresponds to CS model.
One can interprete the above expression for the mass exponent $\theta$ as 
the sum of the fractal dimension $d_f$ and that of its embedding 
space $1$. Using Eq. (\ref{eq:massexponent}) in $\beta=\theta z$ we can further write a generalized scaling relation
\begin{equation} 
\label{eq:massexponent_1}
\beta=(1+d_f)z.
\end{equation}
It is interesting that a similar expression for the exponents $\theta$ and $\beta$ have 
also been found in other phenomena which indicates that these results are
universal in character \cite{ref.fractal_1}. 

To further support our results, we once again use the simple dimensional analysis. According to Eq. (\ref{eq:0}) the 
physical dimension of $c(x,t)$ is $[c]=L^{-(1+d_f)}$ since $[s(t)]=L$ and $\theta=1+d_f$. On the other hand, the 
concentration $c(x,t)$ is defined as 
the number of particles per unit volume of embedding space ($V\sim L^d$ where $d=1$) per unit mass ($M$) and hence
$[c]=L^{-1}M^{-1}$. Now applying the principle of equivalence we obtain
\begin{equation}
\label{eq:masslength}
M(L)\sim L^{d_f}.
\end{equation}
This relation is often regarded as the hallmark for the emergence of fractality.
An object whose mass-length relation satisfies Eq. (\ref{eq:masslength}) with non-integer exponent is said to be
fractal in the sense that if the linear dimension of  the object is increased by a factor of $L$ the mass of the
object is not increased by the same factor. That is, the distribution of mass in the object becomes less dense at 
larger length scale. It proves that the splits of the mass exponent into dimension of the 
fractal ($d_f$) and that of its embedding space ($d=1$) is consistent with the definition of the distribution 
function $c(x,t)$ as well. It is interesting to note that such a simple dimensional analysis can also provide us 
with an answer to the
question: Why is the moment $M_{d_f}=\int_0^\infty x^{d_f}c(x,t)dx$ a conserved quantity?
For an asnwer, we find it conventient to look into the physical dimension of its differential quantity 
$dM_{d_f}=x^{d_f}c(x,t)dx$. Using
the physical dimension $[x]=L$ and $[c(x,t)]=L^{-(1+d_f)}$ in the expression for $dM_{d_f}$, 
we immediately find that it bears no dimension and so is the quantity $M_{d_f}$. Recall that the numerical value
of a dimensionless quantity always remain unchanged upon transition from one unit of measurement
to another within a given class. In the context of the CDA model it implies that the numerical value of $M_{d_f}$ 
remains the same despite the fact that the system size continues to grow with time.
It is due to this reason that we find that the $d_f$th moment of $c(x,t)$ is a conserved quantity.  
We thus see that the simple dimensional analysis proved to be very 
useful in gaining the comprehensive explanations of various results of the CDA model which we have been longing for.

\section{Discussion and Summary}

In this work, we studied the geometric aspect of the condensation-driven aggregation model that
we recently proposed. In the present work we first gave a simple dimensional analysis to the generalized Smoluchowski 
equation as we found it provided not only a deeper insight but also, at the same time, an elegant way 
to look into the problem. 
We then applied the scaling theory and shown that the GS equation admits
simple scaling only if $z(\theta-1)=1$. That is,  
the solution for particle size spectra exhibits transition to dynamic scaling
$c(x,t)\sim t^{-\theta z}\phi(x/t^z)$ with scaling function $\phi(\xi)\sim e^{-\xi}$. 
Substituting the solution for the distribution function $c(x,t)$ into the definition of the $n$th moment 
shows that the moment of order equal to 
${{1}\over{1+2\alpha}}$ is a conserved quantity throughout. In an attempt to know exactly why this value is so special
we invoked the idea of fractal analysis and found that it is in fact the value of the fractal 
dimension of the resulting system. The expression for fractal dimension $d_f={{1}\over{1+2\alpha}}$  
states that as the extent of growth by condensation increases the dimension of the measure or the object decreases
which is quite counter intuitive.

To summarize, we found that the fractal dimension $d_f$ plays 
a pivotal role in describing and understanding the geometric aspect of the CDA model. 
For instance, the dynamics of the system is governed by a conservation law which is the $d_f^{th}$ moment
of the distribution function $c(x,t)$. The exponent of the algebraic growth-law for the mean particle size 
is equal to ${{1}\over{d_f}}$. In terms of $d_f$ we can express the mass exponent 
and the scaling relation in their generalized form such as $\theta=1+d_f$ and $\beta=(1+d_f)z$ respectively. 
A simple dimensional analysis to the distribution function $c(x,t)$ and the use
of $\theta=1+d_f$ led to the well known mass-length relation $M(L)\sim L^{d_f}$. The dimensional analysis also 
revealed that the mass exponent $\theta$ is in fact equal to
the sum of the fractal dimension $d_f$ and that of its space where it is embedded. 
Besides, we have shown that the $d_f$th moment $M_{d_f}$ is actually a dimensionless quantity and we argued that this is 
exactly the reason why $M_{d_f}$ remains time invariant.
We thus see that appreciation of the exponent $d_f$ as the fractal dimension has provided a self-consistent explanation 
to all the results which are found to be independent of the initial particle size distribution. Moreover, we have shown that the results are independent of initial conditions. The ideas developed
in this paper could be taken further by investigating the CDA model for aggregation kernel 
$K(x,y) = (xy)^\omega $. This would be an ideal case in order to study how the onset of 
gelation is modified, if at all, by the presence of growth by heterogeneous condensation. 
We hope to address this issue in our future endeavour.

\end{document}